\documentclass[prd,aps,preprintnumbers,nofootinbib,twocolumn]{revtex4-1}

\usepackage{amscd}
\usepackage{amsmath, bbm} 
\usepackage{amsfonts}
\usepackage{amssymb}
\usepackage{slashed}
\usepackage{multirow}
\usepackage{array}
\usepackage{cancel}
\usepackage{xcolor}
\usepackage{hyperref}
\usepackage{axodraw2}
\usepackage{tikz-cd} 

\begin{document} 
\title{Floccinaucinihilipilification:\\ 
Semisimple extensions of the Standard Model gauge algebra}
\author{B C Allanach} \email{B.C.Allanach@damtp.cam.ac.uk} \affiliation{DAMTP, University of Cambridge, Wilberforce Road, Cambridge, 
CB3 0WA, United Kingdom}
\author{Ben Gripaios} \email{gripaios@hep.phy.cam.ac.uk} \affiliation{Cavendish Laboratory, University of Cambridge, J.J. Thomson
Avenue, Cambridge, CB3 0HE, United Kingdom}
\author{Joseph Tooby-Smith} \email{jss85@cam.ac.uk} \affiliation{Cavendish Laboratory, University of Cambridge, J.J. Thomson
Avenue, Cambridge, CB3 0HE, United Kingdom}
\begin{abstract}
We show how one may classify all semisimple algebras containing the
$\mathfrak{su}(3)\oplus \mathfrak{su}(2) \oplus \mathfrak{u}(1)$ symmetry of
the Standard Model and acting on some given matter sector, enabling theories
beyond the Standard Model with unification (partial or total) of symmetries
({gauge} or global) to be catalogued. With just a single generation of Standard
Model fermions plus a singlet neutrino, the only {gauge} symmetries correspond
to the well-known algebras $\mathfrak{su}(5),\mathfrak{so}(10),$ and
$\mathfrak{su}(4)\oplus \mathfrak{su}(2) \oplus \mathfrak{su}(2)$, but with
two or more generations a limited number of exotic symmetries mixing
flavour, colour, and electroweak degrees of freedom
become possible. We provide a complete
catalogue in the case of 3 generations or fewer and outline how our method 
  generalizes to cases with additional
  matter.
\end{abstract}
\maketitle
%%%%%%%%%%%%%%%%%
\section{Introduction}
%%%%%%%%%%%%%%%%%

In searching for theories of physics beyond the Standard Model (SM), it is of interest to ask how the gauge Lie algebra $\mathfrak{sm}:= \mathfrak{su}(3) \oplus \mathfrak{su}(2) \oplus \mathfrak{u}(1)$ could be extended to a larger Lie algebra $\mathfrak{g} \supset \mathfrak{sm}$. 
To give a useful answer to this question 
requires us to make some plausible assumptions, not least because there are, {\em a priori}, infinitely many such algebras, and because the question is anyway meaningless if we do not specify how $\mathfrak{g}$ acts on the physical degrees of freedom.

{The list of possible $\mathfrak{g}$ becomes not only finite, but also can be
  determined explicitly with the help of a computer, once we specify that
  $\mathfrak{g}$ is semisimple, as is indicated by the fact that ratios of
  hypercharges are simple fractions and by the fact that gauge couplings
  appear to unify, and that $\mathfrak{g}$ acts by a unitary (respectively
  orthogonal) representation on some given complex (respectively real) matter
  fields, so as to preserve the  kinetic terms in the Lagrangian density.} (Strictly speaking, to get a finite list we must discard the largest summand of $\mathfrak{g}$ that acts trivially on the matter, which is anyway of no interest, and identify algebras that lead to equivalent physical theories, as we discuss below. Thus, for us, $\mathfrak{g}$ will act via faithful representation.)
The list may be further curtailed by insisting that $\mathfrak{g}$ be free of {local}
anomalies {(global anomalies require us to specify the gauge group,
  in general, and will not concern us here)} with respect to fermionic matter, so that it can be
gauged.~\footnote{If we discard the requirement that $\mathfrak{g}$ be semisimple, the list becomes infinite, even if we add only a single anomaly-free $\mathfrak{u}(1)$, as Ref.~\cite{Allanach:2020zna} shows.} Such a list can serve as a {\em vade mecum}\/ for model builders.  

In this paper, we find all such $\mathfrak{g}$ in the case where the matter
fields are taken to be the 3 generations of quarks and leptons of the SM
along with 3 $\mathfrak{sm}$-singlet fermions (invoked to give neutrinos their
observed masses), bringing the total number of Weyl fermions to 48. A valid
$\mathfrak{g}$ is then given by an anomaly-free semisimple algebra that contains $\mathfrak{sm}$ and is contained in $\mathfrak{su}(48)$. 
Two such algebras will lead to physical theories that are equivalent if they
are related by an inner automorphism of $\mathfrak{su}(48)$, since such an
automorphism can be effected by a linear change of variables of the fermion
fields, which leaves the path integral invariant.
  They will also lead to
equivalent theories if they are related by an outer automorphism of
$\mathfrak{g}$, since applying such a transformation does not change the image
of $\mathfrak{g}$ in $\mathfrak{su}(48)$. 

Although we study just one example, the methods we use can be generalized at
whim. For example, one could easily include the scalar Higgs fields of the SM
(in which case one seeks all $\mathfrak{g}$ containing $\mathfrak{sm}$ and
contained in $\mathfrak{su} (48)\oplus \mathfrak{so}(4)$) or indeed with $n$
additional fermions and $m$ additional scalars (in which case the containing
algebra is $\mathfrak{su}(48+n)\oplus \mathfrak{so}(4+m)$). In this way, we
end up with a procedure for finding unification symmetries that is so general
that a critic might judge it as being worthless; we therefore feel a fitting name for it is floccinaucinihilipilification.  

To illustrate the results we obtain, it is useful to begin with the simpler
case with just a single generation of quarks and leptons together with a
single $\mathfrak{sm}$-singlet fermion. Here we already know that there are at least 3
possibilities, corresponding to the unification algebras
{$\mathfrak{ps}:=\mathfrak{su}(4)\oplus
\mathfrak{su}(2)^{\oplus 2}$ ({\em i.e.} $\mathfrak{su}(4)\oplus
\mathfrak{su}(2) \oplus \mathfrak{su}(2)$)}~\cite{Pati:1974yy},
$\mathfrak{su}(5)$~\cite{Georgi:1974sy}, and $\mathfrak{so}(10)$~\cite{Fritzsch:1974nn,Georgi:1974my} (which are all
subalgebras of $\mathfrak{su}(16)$; without the extra $\mathfrak{sm}$-singlet, we have just
$\mathfrak{su}(5) \subset \mathfrak{su}(15)$); the `new result' with just a
single generation, then, is that there are, perhaps unsurprisingly, no other
possibilities.~\footnote{In each case there is a single non-trivial outer
  automorphism of $\mathfrak{g}$, so without this equivalence we would find 6
  possibilities, corresponding to the fact that one could assign the fermions
  to carry either of the inequivalent 16-dimensional spinor representations of
  $\mathfrak{so}(10)$, say.}

Things become more interesting with more generations, because of the
possibility of interplay between gauge and flavour symmetries. With two
generations, for example, there are 45 possible algebras, up to
equivalence. Some of these are easily guessed, such as the algebra
$\mathfrak{so}(10) \oplus \mathfrak{su}(2)$, with the right hand
factor acting as a flavour symmetry mixing the 2 generations, along
with $\mathfrak{so}(10)\oplus \mathfrak{su}(5)$, with each summand
acting non-trivially only on a single generation. But there also exist
possibilities that are less easy to guess and which are interesting in
that they mix up flavour and gauge symmetries in {an essential way.} One of these has algebra 
$\mathfrak{su}(8)\oplus \mathfrak{su}(2)^{\oplus 2}$, which
generalizes with $N$ generations to $\mathfrak{su}(4N)\oplus \mathfrak{su}(2)^{\oplus 2}
$. {This construction relies on the obvious embedding $\mathfrak{su}(4)
\oplus \mathfrak{su}(N) \subset \mathfrak{su}(4N)$, showing that it
provides a generalization of the usual Pati-Salam symmetry
$\mathfrak{ps}$ for $N=1$ containing an $\mathfrak{su}(N)$ flavour
symmetry. Thus, whereas in the usual Pati-Salam setup lepton number is interpreted
as the fourth colour, here flavour is to be interpreted as the remaining
$4N-4$ colours!} 

A qualitatively different generalization of the Pati-Salam model with two
generations can be obtained as follows. Since the fermion fields in the
one-generation version form two irreducible representations, there is a
possible $\mathfrak{su}(2)^{\oplus 2}$ flavour symmetry when we
go to two generations, giving the algebra $\mathfrak{su}(4) \oplus
\mathfrak{su}(2)^{\oplus 4} $, with the $32$ fermion fields arranging themselves into the
representation $(\mathbf{4},\mathbf{2},\mathbf{2},\mathbf{1},\mathbf{1}) \oplus
(\overline{\mathbf{4}},\mathbf{1},\mathbf{1},\mathbf{2},\mathbf{2})$. This is
anomaly free because $\mathfrak{su}(2)$ has no anomalous
representations. Noting that $\mathfrak{su}(2)^{\oplus 2}
\cong \mathfrak{so}(4) \subset \mathfrak{so}(5) \cong
\mathfrak{sp}(4)$ and that the defining representation of
$\mathfrak{sp}(4)$ restricts to the $(\mathbf{2},\mathbf{2})$ of
$\mathfrak{su}(2)^{\oplus 2}$, it follows that this can
be enlarged even further to $\mathfrak{su}(4) \oplus \mathfrak{sp}(4)
^{\oplus 2}$, {again leading to an essential mixing of flavour symmetry with
$\mathfrak{sm}$.}

{Since this last construction relies on `accidental' isomorphisms of
low-dimensional Lie algebras, we do not expect it to generalize to $N>2$
generations. Two qualitatively new algebras do appear, however. One
uses the embeddings $\mathfrak{su}(16N) \supset \mathfrak{su}(4)
\oplus \mathfrak{sp}(2N)^{\oplus 2} \subset \mathfrak{su}(4)
\oplus (\mathfrak{so}(N) \oplus \mathfrak{sp}(2))^{\oplus 2} $ along with the
isomorphism $ \mathfrak{sp}(2) \cong  \mathfrak{su}(2)$ to produce an
algebra containing {$\mathfrak{ps}$} along with an $
\mathfrak{so}(N)^{\oplus 2}$ flavour symmetry. The other uses the embeddings 
$\mathfrak{su}(16N) \supset \mathfrak{su}(4)
\oplus \mathfrak{sp}(2N) \oplus \mathfrak{so}(2N) \subset \mathfrak{su}(4)
\oplus \mathfrak{so}(N) \oplus \mathfrak{sp}(2) \oplus \mathfrak{so}(2)
  \oplus \mathfrak{so}(N)$, to produce an algebra containing not {$\mathfrak{ps}$}, but rather its subalgebra $\mathfrak{su}(4)
  \oplus \mathfrak{su}(2) \oplus \mathfrak{so}(2) \supset
  \mathfrak{sm}$, which is not only not semisimple, but is also not
  left-right symmetric. {Again we find a flavour symmetry isomorphic
  to $ \mathfrak{so}(N)^{\oplus 2}$, but now embedded differently in the SM
  flavour symmetry.} These constructions rely on the embeddings
  $\mathfrak{so}({2}) \oplus \mathfrak{so}(N) \subset \mathfrak{so}({2}N)$
  and $\mathfrak{sp}({2}) \oplus \mathfrak{so}(N) \subset
  \mathfrak{sp}({2}N)$. Note that in these examples, flavour symmetry is unified
  with the electroweak symmetry rather than with the strong symmetry, and in a
  variety of ways.~\footnote{{For an even number of generations, we
  also have an embedding $\mathfrak{su}(16N) \supset \mathfrak{su}(4)
  \oplus  \mathfrak{so}(2N)^{\oplus 2} \supset \mathfrak{su}(4) \oplus  (\mathfrak{sp}(2)
  \oplus \mathfrak{sp}(N))^{\oplus 2}$, using the embedding
  $\mathfrak{sp}(2) \oplus \mathfrak{sp}(N) \subset \mathfrak{so}(2N)$.}}}

The upshot is that with 3 generations we get many more algebras
(340, up to equivalence) but all of them can be regarded as variations on the
themes already described. This shows that the model building possibilities are
in fact extremely limited, unless we include additional fermion
fields. Nevertheless, we find a small number of interesting possibilities
which mix gauge and flavour symmetries in an essential way. In particular, if
such symmetries are {gauged}, the corresponding gauge bosons can change both
flavour and colour/electroweak charges of matter fields.

The algebras organize themselves into {24} (respectively {5}) equivalence classes
of semisimple anomaly free algebras that are maximal (respectively minimal)
with respect to inclusion (note that $\mathfrak{su}(48)$ is not anomaly free,
and $\mathfrak{su}(3)\oplus \mathfrak{su}(2) \oplus \mathfrak{u}(1)$ is not
semisimple, so these definitions are cromulent). We list these in
Table~\ref{tab:list_of_puts}. The remaining algebras can be found in
  Table~\ref{tab:list_of_puts_non_max_}. Additional information about the embeddings can be founds in a supplementary file.
%%%%%%%%%%%%%%%%%%%%%%%%%%%%%%%%%%%%%%%%
\section{Theory}
%%%%%%%%%%%%%%%%%%%%%%%%%%%%%%%%%%%%%%%%
We now describe the mathematical formulation of the problem. Because of the need to track automorphisms, this is most easily done using the language of category theory. A suitable category has objects, labelled $(\mathfrak{g},\alpha,\beta,\gamma)$, given by commuting diagrams of the form
\begin{equation} \label{eq:objectincat}
\begin{tikzcd}
& \mathfrak{g}\arrow{rd}{\beta}  &\\
\mathfrak{sm}\arrow{ru}{\alpha} \arrow{rr}{\gamma}& & \mathfrak{su}(48)
\end{tikzcd}
\end{equation}
where $\mathfrak{g}$ is a semisimple Lie algebra, and $\alpha$, $\beta$, $\gamma$ are embeddings, \emph{i.e.} injective maps that preserve Lie brackets. A morphism, labelled $(j,i)$, from $(\mathfrak{g}^\prime,\alpha^\prime,\beta^\prime,\gamma^\prime)$ to $(\mathfrak{g},\alpha,\beta,\gamma)$ is then a commuting diagram of the form
\begin{equation}
\begin{tikzcd}
& \mathfrak{g} \arrow{rd}{\beta}& \\
\mathfrak{sm} \arrow{ru}{\alpha} \arrow[pos=0.2]{rr}{\gamma}& \mathfrak{g}^\prime\arrow{rd}{\beta^\prime}\arrow{u}{j} & \mathfrak{su}(48)\\
\mathfrak{sm}\arrow{ru}{\alpha^\prime} \arrow{u}{\mathrm{id}} \arrow{rr}{\gamma^\prime}& & \mathfrak{su}(48)\arrow{u}{i}
\end{tikzcd}
\end{equation}
where $j$ is an embedding and $i$ is an inner automorphism. We call a morphism $(j,i)$ an equivalence if $j$ is an isomorphism (\emph{i.e.} if $j$ also surjects). We say such a diagram~\ref{eq:objectincat} is maximal (respectively minimal) if the only morphisms out of (respectively into) it are equivalences.

Our goal is then to find all inequivalent diagrams $(\mathfrak{g},\alpha,\beta,\gamma)$ for which $\gamma$ corresponds to the SM embedding. 
To do so, we choose Cartan subalgebras of each algebra appearing in
diagram~\ref{eq:objectincat}, which we denote $\mathfrak{h}_{\mathfrak{sm} }$,
$\mathfrak{h}$ and $\mathfrak{h}_{48}$ for $\mathfrak{sm}$, $\mathfrak{g}$ and
$\mathfrak{su}(48)$ respectively. Up to equivalence, we choose $\alpha$,
$\beta$, and $\gamma$ such that these get embedded into one another
(along with a compatible restriction on morphisms).

Equivalently, using the universal property of the direct sum, we seek a pair of diagrams
\begin{equation} \label{eq:objectinothercat}
\begin{tikzcd}[column sep=2.5pt]
& \mathfrak{g}\arrow{rd}{\beta}  &\\
\mathfrak{su}(3)\oplus \mathfrak{su}(2)\arrow{ru}{\kappa} \arrow{rr}{\rho}& & \mathfrak{su}(48)
\end{tikzcd}
\begin{tikzcd}[column sep=2.5pt]
& \mathfrak{h}\arrow{rd}{\beta |}  &\\
\mathfrak{u}(1)\arrow{ru}{\tilde \kappa} \arrow{rr}{\tilde \rho}& & \mathfrak{h}_{48}
\end{tikzcd}
\end{equation}
where $\gamma=\rho\oplus \tilde \rho$ corresponds to the chosen SM embedding, $\alpha=\kappa\oplus \tilde \kappa$, and $\beta |$ on the right hand side denotes the obvious restriction map.
%%%%%%%%%%%%%%%%%%%%%%%%%%%%%%%%%%%%%%%%
\section{Computation}
%%%%%%%%%%%%%%%%%%%%%%%%%%%%%%%%%%%%%%%% 
{In rough terms, our approach to the computation is as follows. The first step
is to evaluate all embeddings $\beta$ in Eq.~\ref{eq:objectincat}. Up to inner automorphisms of
$\mathfrak{su}(48)$ these are
inequivalent
representations of semisimple Lie algebras of dimension $48$ and so can be
found using standard representation theory techniques. We keep only those
which are anomaly free. For every $\mathfrak{g}$ for which a $\beta$ exists,
we then find all embeddings $\kappa$ {up to outer automorphisms} using the theory of maximal embeddings.  We then find all $\kappa$ and
$\beta$ such that there exists a $\rho$ matching the embedding of the SM up to
inner automorphisms, as in the
left-hand diagram in Eq.~\ref{eq:objectinothercat}. For a given diagram, we
then determine if compatible $\tilde \kappa$ and $\tilde \rho$ exist (taking
account of possible inner-automorphisms which may need to be applied). By
finding all embeddings $j:\mathfrak{g}^\prime \rightarrow \mathfrak{g}$ for
algebras which appear in our final list, {our} program then checks
which are maximal and
which are minimal. }

{The program itself uses projection matrices rather than
  embeddings. Projection matrices describe how the weights project from
 the super-algebra to the sub-algebra. For a given $\mathfrak{g}$, the projection matrices corresponding to potentially allowed $\kappa$'s can be found using the theory
of maximal
embeddings~\cite{Mal44,dyn52:MaximalSubgroups,dynkin1952semisimple,Lorente:1972xw,Yamatsu:2015npn}. (We
use the {\tt Mathematica} program {\tt LieART2.0}~\cite{Feger:2012bs,Feger:2019tvk} to
generate the maximal projection matrices themselves). One has to take care to
ensure that these projection matrices lift correctly to embeddings, and to ensure that they define a unique diagram~\ref{eq:objectincat} up to equivalence.}

The output of the program (provided in a supplementary file) consists of the
highest weights of the representation specified by $\beta$ and the projection
matrix of the embedding $\kappa$, to which we have appended a final row
specifying $\tilde \kappa$ (to wit, acting on the weights of $\mathfrak{g}$,
this row returns the corresponding $\mathfrak{u}(1) \subset \mathfrak{sm}$
charges).

\renewcommand{\arraystretch}{1.5} 
\begin{table*}
\fontsize{8}{9.2}\selectfont
\begin{tabular}{|c|r|l|}
\hline  \multicolumn{3}{|c|}{\small Maximal} \\ \hline
& {\small Algebra}&\small Fermion representations corresponding to $\beta$ \\ \hline 1 &  $\mathfrak{so}(10) \oplus \mathfrak{su}(2)$ & $({\mathbf{16}}, {\mathbf{3}})$
 %%%%%%%%%%%%%%%%%%%%%%%%%%%%%%%% 
  \\ \hline 2 &  $\mathfrak{so}(10)^{\oplus 3} $ & $({\mathbf{16}}, {\mathbf{1}}, {\mathbf{1}}) \oplus ({\mathbf{1}}, {\mathbf{16}}, {\mathbf{1}}) \oplus ({\mathbf{1}}, {\mathbf{1}}, {\mathbf{16}})$
 %%%%%%%%%%%%%%%%%%%%%%%%%%%%%%%% 
  \\ \hline 3 &  $\mathfrak{so}(10)^{\oplus 2}  \oplus \mathfrak{su}(2)$ & $({\mathbf{16}}, {\mathbf{1}}, {\mathbf{1}}) \oplus ({\mathbf{1}}, {\mathbf{16}}, {\mathbf{2}})$
 %%%%%%%%%%%%%%%%%%%%%%%%%%%%%%%% 
  \\ \hline 4 &  $\mathfrak{su}(4) \oplus \mathfrak{sp}(6)^{\oplus 2} $ & $(\overline{\mathbf{4}}, {\mathbf{6}}, {\mathbf{1}}) \oplus ({\mathbf{4}}, {\mathbf{1}}, {\mathbf{6}})$
 %%%%%%%%%%%%%%%%%%%%%%%%%%%%%%%% 
  \\ \hline 5 &  $\mathfrak{su}(4)^{\oplus 2}  \oplus \mathfrak{sp}(6)$ & $(\overline{\mathbf{4}}, {\mathbf{6}}, {\mathbf{1}}) \oplus ({\mathbf{4}}, {\mathbf{1}}, {\mathbf{6}})$
 %%%%%%%%%%%%%%%%%%%%%%%%%%%%%%%% 
  \\ \hline 6 &  $\mathfrak{su}(12) \oplus \mathfrak{su}(2)^{\oplus 2} $ & $(\overline{\mathbf{12}}, {\mathbf{2}}, {\mathbf{1}}) \oplus ({\mathbf{12}}, {\mathbf{1}}, {\mathbf{2}})$
 %%%%%%%%%%%%%%%%%%%%%%%%%%%%%%%% 
  \\ \hline 7 &  $\mathfrak{su}(4) \oplus \mathfrak{sp}(4)^{\oplus 2}  \oplus \mathfrak{so}(10)$ & $(\overline{\mathbf{4}}, {\mathbf{4}}, {\mathbf{1}}, {\mathbf{1}}) \oplus ({\mathbf{4}}, {\mathbf{1}}, {\mathbf{4}}, {\mathbf{1}}) \oplus ({\mathbf{1}}, {\mathbf{1}}, {\mathbf{1}}, {\mathbf{16}})$
 %%%%%%%%%%%%%%%%%%%%%%%%%%%%%%%% 
  \\ \hline 8 &  $\mathfrak{su}(5) \oplus \mathfrak{su}(2)^{\oplus 3} $ & $(\overline{\mathbf{5}}, {\mathbf{3}}, {\mathbf{1}}, {\mathbf{1}}) \oplus ({\mathbf{10}}, {\mathbf{1}}, {\mathbf{3}}, {\mathbf{1}}) \oplus ({\mathbf{1}}, {\mathbf{1}}, {\mathbf{1}}, {\mathbf{2}}) \oplus (\mathbf{1},\mathbf{1},\mathbf{1},\mathbf{1})$
 %%%%%%%%%%%%%%%%%%%%%%%%%%%%%%%% 
  \\ \hline 9 &  $\mathfrak{su}(5) \oplus \mathfrak{su}(2)^{\oplus 3} $ & $(\overline{\mathbf{5}}, {\mathbf{3}}, {\mathbf{1}}, {\mathbf{1}}) \oplus ({\mathbf{10}}, {\mathbf{1}}, {\mathbf{3}}, {\mathbf{1}}) \oplus ({\mathbf{1}}, {\mathbf{1}}, {\mathbf{1}}, {\mathbf{3}})$
 %%%%%%%%%%%%%%%%%%%%%%%%%%%%%%%% 
  \\ \hline 10 &  $\mathfrak{su}(5) \oplus \mathfrak{su}(2)^{\oplus 3} $ & $(\overline{\mathbf{5}}, {\mathbf{1}}, {\mathbf{1}}, {\mathbf{1}}) \oplus (\overline{\mathbf{5}}, {\mathbf{2}}, {\mathbf{1}}, {\mathbf{1}}) \oplus ({\mathbf{10}}, {\mathbf{1}}, {\mathbf{3}}, {\mathbf{1}}) \oplus ({\mathbf{1}}, {\mathbf{1}}, {\mathbf{1}}, {\mathbf{2}}) \oplus (\mathbf{1},\mathbf{1},\mathbf{1},\mathbf{1})$
 %%%%%%%%%%%%%%%%%%%%%%%%%%%%%%%% 
  \\ \hline 11 &  $\mathfrak{su}(5) \oplus \mathfrak{su}(2)^{\oplus 3} $ & $(\overline{\mathbf{5}}, {\mathbf{1}}, {\mathbf{1}}, {\mathbf{1}}) \oplus (\overline{\mathbf{5}}, {\mathbf{2}}, {\mathbf{1}}, {\mathbf{1}}) \oplus ({\mathbf{10}}, {\mathbf{1}}, {\mathbf{3}}, {\mathbf{1}}) \oplus ({\mathbf{1}}, {\mathbf{1}}, {\mathbf{1}}, {\mathbf{3}})$
 %%%%%%%%%%%%%%%%%%%%%%%%%%%%%%%% 
  \\ \hline 12 &  $\mathfrak{su}(5) \oplus \mathfrak{su}(2)^{\oplus 3} $ & $({\mathbf{10}}, {\mathbf{1}}, {\mathbf{1}}, {\mathbf{1}}) \oplus (\overline{\mathbf{5}}, {\mathbf{3}}, {\mathbf{1}}, {\mathbf{1}}) \oplus ({\mathbf{10}}, {\mathbf{1}}, {\mathbf{2}}, {\mathbf{1}}) \oplus ({\mathbf{1}}, {\mathbf{1}}, {\mathbf{1}}, {\mathbf{2}}) \oplus (\mathbf{1},\mathbf{1},\mathbf{1},\mathbf{1})$
 %%%%%%%%%%%%%%%%%%%%%%%%%%%%%%%% 
  \\ \hline 13 &  $\mathfrak{su}(5) \oplus \mathfrak{su}(2)^{\oplus 3} $ & $({\mathbf{10}}, {\mathbf{1}}, {\mathbf{1}}, {\mathbf{1}}) \oplus (\overline{\mathbf{5}}, {\mathbf{3}}, {\mathbf{1}}, {\mathbf{1}}) \oplus ({\mathbf{10}}, {\mathbf{1}}, {\mathbf{2}}, {\mathbf{1}}) \oplus ({\mathbf{1}}, {\mathbf{1}}, {\mathbf{1}}, {\mathbf{3}})$
 %%%%%%%%%%%%%%%%%%%%%%%%%%%%%%%% 
  \\ \hline 14 &  $\mathfrak{su}(5)^{\oplus 2}  \oplus \mathfrak{so}(10) \oplus \mathfrak{su}(2)$ & $(\overline{\mathbf{5}}, {\mathbf{1}}, {\mathbf{1}}, {\mathbf{1}}) \oplus ({\mathbf{10}}, {\mathbf{1}}, {\mathbf{1}}, {\mathbf{1}}) \oplus ({\mathbf{1}}, \overline{\mathbf{5}}, {\mathbf{1}}, {\mathbf{1}}) \oplus ({\mathbf{1}}, {\mathbf{10}}, {\mathbf{1}}, {\mathbf{1}}) \oplus ({\mathbf{1}}, {\mathbf{1}}, {\mathbf{16}}, {\mathbf{1}}) \oplus ({\mathbf{1}}, {\mathbf{1}}, {\mathbf{1}}, {\mathbf{2}})$
 %%%%%%%%%%%%%%%%%%%%%%%%%%%%%%%% 
  \\ \hline 15 &  $\mathfrak{su}(5)^{\oplus 3}  \oplus \mathfrak{su}(2)$ & $(\overline{\mathbf{5}}, {\mathbf{1}}, {\mathbf{1}}, {\mathbf{1}}) \oplus ({\mathbf{10}}, {\mathbf{1}}, {\mathbf{1}}, {\mathbf{1}}) \oplus ({\mathbf{1}}, \overline{\mathbf{5}}, {\mathbf{1}}, {\mathbf{1}}) \oplus ({\mathbf{1}}, {\mathbf{10}}, {\mathbf{1}}, {\mathbf{1}}) \oplus ({\mathbf{1}}, {\mathbf{1}}, \overline{\mathbf{5}}, {\mathbf{1}}) \oplus ({\mathbf{1}}, {\mathbf{1}}, {\mathbf{10}}, {\mathbf{1}}) \oplus ({\mathbf{1}}, {\mathbf{1}}, {\mathbf{1}}, {\mathbf{3}})$
 %%%%%%%%%%%%%%%%%%%%%%%%%%%%%%%% 
  \\ \hline 16 &  $\mathfrak{su}(8) \oplus \mathfrak{so}(10) \oplus \mathfrak{su}(2)^{\oplus 2} $ & $({\mathbf{1}}, {\mathbf{16}}, {\mathbf{1}}, {\mathbf{1}}) \oplus (\overline{\mathbf{8}}, {\mathbf{1}}, {\mathbf{2}}, {\mathbf{1}}) \oplus ({\mathbf{8}}, {\mathbf{1}}, {\mathbf{1}}, {\mathbf{2}})$
 %%%%%%%%%%%%%%%%%%%%%%%%%%%%%%%% 
  \\ \hline 17 &  $\mathfrak{su}(4) \oplus \mathfrak{sp}(4) \oplus \mathfrak{so}(10) \oplus \mathfrak{su}(2)^{\oplus 2} $ & $(\overline{\mathbf{4}}, {\mathbf{4}}, {\mathbf{1}}, {\mathbf{1}}, {\mathbf{1}}) \oplus ({\mathbf{1}}, {\mathbf{1}}, {\mathbf{16}}, {\mathbf{1}}, {\mathbf{1}}) \oplus ({\mathbf{4}}, {\mathbf{1}}, {\mathbf{1}}, {\mathbf{2}}, {\mathbf{2}})$
 %%%%%%%%%%%%%%%%%%%%%%%%%%%%%%%% 
  \\ \hline 18 &  $\mathfrak{su}(4) \oplus \mathfrak{sp}(4) \oplus \mathfrak{so}(10) \oplus \mathfrak{su}(2)^{\oplus 2} $ & $(\overline{\mathbf{4}}, {\mathbf{4}}, {\mathbf{1}}, {\mathbf{1}}, {\mathbf{1}}) \oplus ({\mathbf{1}}, {\mathbf{1}}, {\mathbf{16}}, {\mathbf{1}}, {\mathbf{1}}) \oplus ({\mathbf{4}}, {\mathbf{1}}, {\mathbf{1}}, {\mathbf{2}}, {\mathbf{2}})$
 %%%%%%%%%%%%%%%%%%%%%%%%%%%%%%%% 
  \\ \hline 19 &  $\mathfrak{su}(4) \oplus \mathfrak{sp}(6) \oplus \mathfrak{su}(2)^{\oplus 3} $ & $(\overline{\mathbf{4}}, {\mathbf{6}}, {\mathbf{1}}, {\mathbf{1}}, {\mathbf{1}}) \oplus ({\mathbf{4}}, {\mathbf{1}}, {\mathbf{2}}, {\mathbf{2}}, {\mathbf{1}}) \oplus ({\mathbf{4}}, {\mathbf{1}}, {\mathbf{1}}, {\mathbf{1}}, {\mathbf{2}})$
 %%%%%%%%%%%%%%%%%%%%%%%%%%%%%%%% 
  \\ \hline 20 &  $\mathfrak{su}(4) \oplus \mathfrak{sp}(6) \oplus \mathfrak{su}(2)^{\oplus 3} $ & $(\overline{\mathbf{4}}, {\mathbf{6}}, {\mathbf{1}}, {\mathbf{1}}, {\mathbf{1}}) \oplus ({\mathbf{4}}, {\mathbf{1}}, {\mathbf{2}}, {\mathbf{2}}, {\mathbf{1}}) \oplus ({\mathbf{4}}, {\mathbf{1}}, {\mathbf{1}}, {\mathbf{1}}, {\mathbf{2}})$
 %%%%%%%%%%%%%%%%%%%%%%%%%%%%%%%% 
  \\ \hline 21 &  $\mathfrak{su}(4)^{\oplus 2}  \oplus \mathfrak{su}(2)^{\oplus 3} $ & $(\overline{\mathbf{4}}, {\mathbf{6}}, {\mathbf{1}}, {\mathbf{1}}, {\mathbf{1}}) \oplus ({\mathbf{4}}, {\mathbf{1}}, {\mathbf{2}}, {\mathbf{2}}, {\mathbf{1}}) \oplus ({\mathbf{4}}, {\mathbf{1}}, {\mathbf{1}}, {\mathbf{1}}, {\mathbf{2}})$
 %%%%%%%%%%%%%%%%%%%%%%%%%%%%%%%% 
  \\ \hline 22 &  $\mathfrak{su}(5) \oplus \mathfrak{so}(10) \oplus \mathfrak{su}(2)^{\oplus 3} $ & $({\mathbf{1}}, {\mathbf{16}}, {\mathbf{1}}, {\mathbf{1}}, {\mathbf{1}}) \oplus (\overline{\mathbf{5}}, {\mathbf{1}}, {\mathbf{2}}, {\mathbf{1}}, {\mathbf{1}}) \oplus ({\mathbf{10}}, {\mathbf{1}}, {\mathbf{1}}, {\mathbf{2}}, {\mathbf{1}}) \oplus ({\mathbf{1}}, {\mathbf{1}}, {\mathbf{1}}, {\mathbf{1}}, {\mathbf{2}})$
 %%%%%%%%%%%%%%%%%%%%%%%%%%%%%%%% 
  \\ \hline 23 &  $\mathfrak{su}(5)^{\oplus 2}  \oplus \mathfrak{su}(2)^{\oplus 3} $ & $({\mathbf{1}}, \overline{\mathbf{5}}, {\mathbf{1}}, {\mathbf{1}}, {\mathbf{1}}) \oplus ({\mathbf{1}}, {\mathbf{10}}, {\mathbf{1}}, {\mathbf{1}}, {\mathbf{1}}) \oplus (\overline{\mathbf{5}}, {\mathbf{1}}, {\mathbf{2}}, {\mathbf{1}}, {\mathbf{1}}) \oplus ({\mathbf{10}}, {\mathbf{1}}, {\mathbf{1}}, {\mathbf{2}}, {\mathbf{1}}) \oplus ({\mathbf{1}}, {\mathbf{1}}, {\mathbf{1}}, {\mathbf{1}}, {\mathbf{3}})$
 %%%%%%%%%%%%%%%%%%%%%%%%%%%%%%%% 
  \\ \hline 24 &  $\mathfrak{su}(4) \oplus \mathfrak{so}(10) \oplus \mathfrak{su}(2)^{\oplus 4} $ & $({\mathbf{1}}, {\mathbf{16}}, {\mathbf{1}}, {\mathbf{1}}, {\mathbf{1}}, {\mathbf{1}}) \oplus ({\mathbf{4}}, {\mathbf{1}}, {\mathbf{2}}, {\mathbf{2}}, {\mathbf{1}}, {\mathbf{1}}) \oplus (\overline{\mathbf{4}}, {\mathbf{1}}, {\mathbf{1}}, {\mathbf{1}}, {\mathbf{2}}, {\mathbf{2}})$
 %%%%%%%%%%%%%%%%%%%%%%%%%%%%%%%% 
  \\ \hline   \multicolumn{3}{|c|}{\small Minimal} \\ \hline 25 &  $\mathfrak{su}(5)$ & $({\overline{\mathbf{5}}})^{\oplus 3}  \oplus ({{\mathbf{10}}})^{\oplus 3}  \oplus (\mathbf{1})^{\oplus 3}$
 %%%%%%%%%%%%%%%%%%%%%%%%%%%%%%%% 
  \\ \hline 26 &  $\mathfrak{su}(4) \oplus \mathfrak{su}(2)^{\oplus 2} $ & $(\overline{\mathbf{4}}, {\mathbf{2}}, {\mathbf{1}})^{\oplus 3}  \oplus ({\mathbf{4}}, {\mathbf{1}}, {\mathbf{2}})^{\oplus 3} $
 %%%%%%%%%%%%%%%%%%%%%%%%%%%%%%%% 
  \\ \hline 27 &  $\mathfrak{su}(4)^{\oplus 2}  \oplus \mathfrak{su}(2)$ & $(\overline{\mathbf{4}}, {\mathbf{6}}, {\mathbf{1}}) \oplus ({\mathbf{4}}, {\mathbf{1}}, {\mathbf{2}})^{\oplus 3} $
 %%%%%%%%%%%%%%%%%%%%%%%%%%%%%%%% 
  \\ \hline 28 &  $\mathfrak{su}(4) \oplus \mathfrak{su}(5) \oplus \mathfrak{su}(2)^{\oplus 2} $ & $({\mathbf{1}}, \overline{\mathbf{5}}, {\mathbf{1}}, {\mathbf{1}})^{\oplus 2}  \oplus ({\mathbf{1}}, {\mathbf{10}}, {\mathbf{1}}, {\mathbf{1}})^{\oplus 2}  \oplus (\overline{\mathbf{4}}, {\mathbf{1}}, {\mathbf{2}}, {\mathbf{1}}) \oplus ({\mathbf{4}}, {\mathbf{1}}, {\mathbf{1}}, {\mathbf{2}}) \oplus (\mathbf{1},\mathbf{1},\mathbf{1},\mathbf{1})^{\oplus 2}$
 %%%%%%%%%%%%%%%%%%%%%%%%%%%%%%%% 
  \\ \hline 29 &  $\mathfrak{su}(4) \oplus \mathfrak{su}(5) \oplus \mathfrak{su}(2)^{\oplus 2} $ & $({\mathbf{1}}, \overline{\mathbf{5}}, {\mathbf{1}}, {\mathbf{1}}) \oplus ({\mathbf{1}}, {\mathbf{10}}, {\mathbf{1}}, {\mathbf{1}}) \oplus (\overline{\mathbf{4}}, {\mathbf{1}}, {\mathbf{2}}, {\mathbf{1}})^{\oplus 2}  \oplus ({\mathbf{4}}, {\mathbf{1}}, {\mathbf{1}}, {\mathbf{2}})^{\oplus 2}  \oplus (\mathbf{1},\mathbf{1},\mathbf{1},\mathbf{1})$
 %%%%%%%%%%%%%%%%%%%%%%%%%%%%%%%% 
  \\ \hline 
  \end{tabular}
\caption{\label{tab:list_of_puts} All maximal and minimal anomaly-free
  algebras for exactly $3$ generations of SM fermions plus 3 right-handed neutrinos.}
\end{table*}

This approach results in a number of practical issues when it comes to carrying out the computation, which we now describe, along with their resolutions, in rough order of importance. 

{\em (i)}\/ Since $\mathfrak{su}(2)$ has anomaly-free irreducible
representations of every dimension, there are many possible anomaly-free
embeddings of ideals of $\mathfrak{g}$ made up of $\mathfrak{su}(2)$s in
$\mathfrak{su}(48)$.  
For example, there are ${\mathcal O}(10^5)$ for $\mathfrak{g}=\mathfrak{su}(2)$
and ${\mathcal   O}(10^7)$ 
for $\mathfrak{g}=\mathfrak{su}(2)^{\oplus 2}$. We reduce this by
first finding all possible $\beta$ for $\mathfrak{g}$ without an
$\mathfrak{su}(2)$ ideal and then requiring that they contain
$\mathfrak{su}(3) \subset \mathfrak{sm}$  
(here we use the fact that the restriction
of $\kappa$ to $\mathfrak{su}(3)$ has to map trivially into any $\mathfrak{su}(2)$ ideal of $\mathfrak{g}$). To these $\mathfrak{g}$ we add all possible ideals made up of $\mathfrak{su}(2)$s and retest to see if a full $\kappa$ exists.

{\em (ii)}\/ Even after ignoring ideals made up of $\mathfrak{su}(2)$s, there
are still ${\mathcal O}(10^6)$ anomaly free embeddings of $\mathfrak{g}$. We
determine these in a bottom-up fashion by first finding the ${\mathcal O} (10^3)$
anomaly-free representations of dimension 48 of the ${\mathcal O}(10^2)$
possible simple $\mathfrak{g}$ (\emph{e.g.} $\mathbf{5}\oplus
\overline{\mathbf{10}}$ plus 33 singlets of $\mathfrak{su}(5)$) and then using
these to find all anomaly-free representations of possible  semisimple
$\mathfrak{g}$ of dimension 48. Here, we use the fact that a representation of a
semisimple algebra is anomaly-free iff.\ its restriction to any simple ideal
is anomaly-free.  For example, the representations of $\mathfrak{su}(4) \oplus
\mathfrak{sp}(4)$ 
given by
\begin{enumerate}
\item $(\mathbf{4},\mathbf{4})\oplus (\overline{\mathbf{4}},\mathbf{1})^{\oplus 4}{\oplus(\mathbf{1},\mathbf{1})^{\oplus 16}}$
\item$(\overline{\mathbf{4}},\mathbf{4})\oplus (\mathbf{4},\mathbf{1})^{\oplus 4}{\oplus(\mathbf{1},\mathbf{1})^{\oplus 16}}$
\item $(\mathbf{4},\mathbf{1})^{\oplus 4}\oplus (\overline{\mathbf{4}},\mathbf{1})^{\oplus 4}\oplus (\mathbf{1},\mathbf{4})^{\oplus 4}$
\end{enumerate}
are anomaly-free because they all reduce to 
the $\mathbf{4}^{\oplus 4}\oplus \overline{\mathbf{4}}^{\oplus 4}$ of  $\mathfrak{su}(4)$ and the $\mathbf{4}^{\oplus 4}$ of $\mathfrak{sp}(4)$ (plus the appropriate number of singlets). This bottom-up method is also used later to find all representations when ideals made up of $\mathfrak{su}(2)$s are included.  

{\em (iii)}\/ Finding the possible representations of semisimple algebras
above requires considering a large number of permutations of a list (as do
other steps in the calculation, {\em e.g.}\/ finding $\kappa$ from maximal
embeddings). For example, to find the anomaly-free representations 1-3 above requires consideration of around
$500$ different permutations. 
The computation is greatly expedited by the
use of an algorithm that determines which permutations
can be skipped based on previous cases.

Two more minor improvements are as follows: {\em (iv)}\/ the fact that one can discard $\beta$ whose corresponding representation
has a non-trivial part of dimension fewer than $45$ (or $36$ before we include $\mathfrak{su}(2)$ ideals), since these cannot lead to a valid $\alpha$;
{\em (v)}\/ in a similar vein, no $\alpha$ exists for those $\beta$ whose corresponding representations have more than $3$ vanishing weights, or weights in negative pairs, since such weights must be associated with $\mathfrak{sm}$-singlets of which there are just 3.

The program took less than an hour to run on a personal computer. As such, model-builders should find it easy to adapt it to other cases of interest.~\footnote{The programs and instructions on their
  use can be
  found in the ancillary information attached to the {\tt arXiv} preprint
  version of the present paper.}

\section{Closing Remarks}
{We have produced, for the first time, a definitive list of
  semisimple Lie
    algebras that contain the SM Lie algebra, are free of local
    anomalies, and 
  act by a unitary representation on SM fermions plus 3 singlet
  neutrinos.} Such extensions can mix
  {flavour, colour, and electroweak symmetries in non-trivial ways.}
  {Table~\ref{tab:list_of_puts} shows the {24} maximal algebras and {5}
    minimal ones, whereas the remainder are provided in Table~\ref{tab:list_of_puts_non_max_}.
  In total, there are 340 physically-inequivalent algebras.}
  No exceptional Lie algebras appear, since they require
  fermions beyond those in the SM plus 3 singlet neutrinos.
  Many of the symmetries {listed} are semi-familiar, being variations on the theme of
  well-known unification and flavour symmetries. 
  A few of the symmetries in our catalogue have the novel feature of
  combining unification and flavour symmetries in an essential way, motivating
  their further study. For
  example, we have
  $\mathfrak{su}(12)
  \oplus \mathfrak{su}(2)^{\oplus 2}$,~\footnote{The $\mathfrak{su}(12)
  \oplus \mathfrak{su}(2) \oplus \mathfrak{su}(2)$ example appeared in a
  machine-learning scan of a
  subset of Type IIA orientifolds on $T^6/(\mathbb{Z}_2 \times \mathbb{Z}_2)$
  with intersecting D6-branes~\cite{Li:2019miw}}
  {$\mathfrak{su}(8)\oplus \mathfrak{su}(2)^{\oplus 2} $},
 {$\mathfrak{su}(4) \oplus \mathfrak{sp}(6)^{\oplus 2}$,\footnote{{The $\mathfrak{su}(4) \oplus \mathfrak{sp}(6)^{\oplus 2}$ example previously appeared in the context of the generation problem in~\cite{Kuo:1984md} (see also~\cite{Kuo:1984gz}).}}
  $\mathfrak{su}(4) \oplus \mathfrak{sp}(6) \oplus \mathfrak{so}(6)$,}
  and $\mathfrak{su}(4) \oplus \mathfrak{sp}(4)^{\oplus 2}$.  
 
  {By concentrating upon semisimple algebras, we have neglected the
  class of models with abelian summands in their gauge Lie algebras. Previously,
  a few very particular cases with additional summands of $\mathfrak{u}(1)$ have had their
  anomaly cancellation conditions solved
  with quite some effort~\cite{Costa:2019zzy,Allanach:2019gwp,Costa:2020krs,Allanach:2020zna,Costa:2020dph,Dobrescu:2020evn} but
  there are currently no known methods for a fixed matter field content, but
  general Lie algebra.
  Including such abelian summands then implies a significant increase in
  difficulty as compared to not including them. One could imagine having to
  find particular geometric constructions for each possibility of the
  semisimple algebra.
  In order to sidestep this difficulty, we have narrowed the scope to examine
  only semisimple extensions. This, however, implicitly includes all cases
  where  
  an abelian summand is ultimately derived from a semisimple one through its
  spontaneous breaking, which could happen at an intermediate stage between
  $\mathfrak{g}$ and $\mathfrak{sm}$.}
  
{Adding additional matter fields changes the list, but is
straightforward to carry out, in principle. An interesting example to investigate
would be to add a Dirac fermion in the same representation as the
Higgs boson, since it constitutes a
  viable candidate for weakly-interacting thermal relic
  dark matter.}

%%%%%%%%%%%%%%%%%%%%%%
\section{Acknowldegments}
%%%%%%%%%%%%%%%%%%%%%%
{We thank J. Davighi and} other members of the Cambridge Pheno Working Group for
discussions. This work was supported by STFC consolidated
grants ST/P000681/1, ST/T000694/1, and ST/S505316/1.
\bibliography{flocc}

\renewcommand{\arraystretch}{1.5} 
\setcounter{table}{1}
\makeatletter\onecolumngrid@push\makeatother\begin{table*}
\fontsize{8}{9.2}\selectfont
\begin{tabular}{|c|r|l|} \hline
\multicolumn{3}{|c|}{\small Non-Maximal and Non-Minimal algebras} \\ \hline
& {\small Algebra}&\small Fermion representations corresponding to \
$\beta$ \\ \hline 
30 &  $\mathfrak{so}(10)$ & $({\mathbf{16}})^{\oplus 3}$
 %%%%%%%%%%%%%%%%%%%%%%%%%%%%%%%% 
  \\ \hline 31 &  $\mathfrak{so}(10)^{\oplus 2}$ & $({\mathbf{16}}, {\mathbf{1}}) \oplus ({\mathbf{1}}, {\mathbf{16}})^{\oplus 2}$
 %%%%%%%%%%%%%%%%%%%%%%%%%%%%%%%% 
  \\ \hline 32 &  $\mathfrak{so}(10) \oplus \mathfrak{su}(2)$ & $({\mathbf{16}}, {\mathbf{1}}) \oplus ({\mathbf{16}}, {\mathbf{2}})$
 %%%%%%%%%%%%%%%%%%%%%%%%%%%%%%%% 
  \\ \hline 33 &  $\mathfrak{su}(5) \oplus \mathfrak{so}(10)$ & $(\overline{\mathbf{5}}, {\mathbf{1}}) \oplus ({\mathbf{10}}, {\mathbf{1}}) \oplus ({\mathbf{1}}, {\mathbf{16}})^{\oplus 2} \oplus ({\mathbf{1}}, {\mathbf{1}})$
 %%%%%%%%%%%%%%%%%%%%%%%%%%%%%%%% 
  \\ \hline 34 &  $\mathfrak{su}(5) \oplus \mathfrak{so}(10)$ & $(\overline{\mathbf{5}}, {\mathbf{1}})^{\oplus 2} \oplus ({\mathbf{10}}, {\mathbf{1}})^{\oplus 2} \oplus ({\mathbf{1}}, {\mathbf{16}}) \oplus ({\mathbf{1}}, {\mathbf{1}})^{\oplus 2}$
 %%%%%%%%%%%%%%%%%%%%%%%%%%%%%%%% 
  \\ \hline 35 &  $\mathfrak{su}(5) \oplus \mathfrak{su}(2)$ & $(\overline{\mathbf{5}}, {\mathbf{3}}) \oplus ({\mathbf{10}}, {\mathbf{3}}) \oplus ({\mathbf{1}}, {\mathbf{1}})^{\oplus 3}$
 %%%%%%%%%%%%%%%%%%%%%%%%%%%%%%%% 
  \\
\hline 
\end{tabular}
\caption{\label{tab:list_of_puts_non_max_} All algebras for exactly $3$ generations of SM fermions plus 3 right-handed neutrinos which are neither maximal nor minimal.}
\end{table*}

\renewcommand{\arraystretch}{1.5} 
\setcounter{table}{1}
\makeatletter\onecolumngrid@push\makeatother\begin{table*}
\fontsize{8}{9.2}\selectfont
\begin{tabular}{|c|r|l|} \hline
\multicolumn{3}{|c|}{\small Non-Maximal and Non-Minimal algebras} \\ \hline
& {\small Algebra}&\small Fermion representations corresponding to \
$\beta$ \\ \hline 
73 &  $\mathfrak{su}(5) \oplus \mathfrak{su}(2)^{\oplus 2}$ & $(\overline{\mathbf{5}}, {\mathbf{3}}, {\mathbf{1}}) \oplus ({\mathbf{10}}, {\mathbf{1}}, {\mathbf{3}}) \oplus ({\mathbf{1}}, {\mathbf{1}}, {\mathbf{1}})^{\oplus 3}$
 %%%%%%%%%%%%%%%%%%%%%%%%%%%%%%%% 
  \\ \hline 74 &  $\mathfrak{su}(5) \oplus \mathfrak{su}(2)^{\oplus 2}$ & $(\overline{\mathbf{5}}, {\mathbf{1}}, {\mathbf{1}}) \oplus (\overline{\mathbf{5}}, {\mathbf{2}}, {\mathbf{1}}) \oplus ({\mathbf{10}}, {\mathbf{1}}, {\mathbf{3}}) \oplus ({\mathbf{1}}, {\mathbf{1}}, {\mathbf{1}})^{\oplus 3}$
 %%%%%%%%%%%%%%%%%%%%%%%%%%%%%%%% 
  \\ \hline 75 &  $\mathfrak{su}(5) \oplus \mathfrak{su}(2)^{\oplus 2}$ & $(\overline{\mathbf{5}}, {\mathbf{3}}, {\mathbf{1}}) \oplus ({\mathbf{10}}, {\mathbf{1}}, {\mathbf{3}}) \oplus ({\mathbf{1}}, {\mathbf{1}}, {\mathbf{2}}) \oplus ({\mathbf{1}}, {\mathbf{1}}, {\mathbf{1}})$
 %%%%%%%%%%%%%%%%%%%%%%%%%%%%%%%% 
  \\ \hline 76 &  $\mathfrak{su}(5) \oplus \mathfrak{su}(2)^{\oplus 2}$ & $(\overline{\mathbf{5}}, {\mathbf{3}}, {\mathbf{1}}) \oplus ({\mathbf{10}}, {\mathbf{1}}, {\mathbf{3}}) \oplus ({\mathbf{1}}, {\mathbf{1}}, {\mathbf{3}})$
 %%%%%%%%%%%%%%%%%%%%%%%%%%%%%%%% 
  \\ \hline 77 &  $\mathfrak{su}(5) \oplus \mathfrak{su}(2)^{\oplus 2}$ & $(\overline{\mathbf{5}}, {\mathbf{3}}, {\mathbf{1}}) \oplus ({\mathbf{10}}, {\mathbf{1}}, {\mathbf{3}}) \oplus ({\mathbf{1}}, {\mathbf{2}}, {\mathbf{1}}) \oplus ({\mathbf{1}}, {\mathbf{1}}, {\mathbf{1}})$
 %%%%%%%%%%%%%%%%%%%%%%%%%%%%%%%% 
  \\ \hline 78 &  $\mathfrak{su}(5) \oplus \mathfrak{su}(2)^{\oplus 2}$ & $(\overline{\mathbf{5}}, {\mathbf{3}}, {\mathbf{1}}) \oplus ({\mathbf{10}}, {\mathbf{1}}, {\mathbf{3}}) \oplus ({\mathbf{1}}, {\mathbf{3}}, {\mathbf{1}})$
 %%%%%%%%%%%%%%%%%%%%%%%%%%%%%%%% 
  \\
\hline 
\end{tabular}
\caption{Continued}
\makeatletter\onecolumngrid@pop\makeatother\end{table*}

\renewcommand{\arraystretch}{1.5} 
\setcounter{table}{1}
\makeatletter\onecolumngrid@push\makeatother\begin{table*}
\fontsize{8}{9.2}\selectfont
\begin{tabular}{|c|r|l|} \hline
\multicolumn{3}{|c|}{\small Non-Maximal and Non-Minimal algebras} \\ \hline
& {\small Algebra}&\small Fermion representations corresponding to \
$\beta$ \\ \hline 
117 &  $\mathfrak{su}(5)^{\oplus 2} \oplus \mathfrak{su}(2)$ & $({\mathbf{10}}, {\mathbf{1}}, {\mathbf{1}})^{\oplus 2} \oplus ({\mathbf{1}}, \overline{\mathbf{5}}, {\mathbf{1}}) \oplus ({\mathbf{1}}, {\mathbf{10}}, {\mathbf{1}}) \oplus (\overline{\mathbf{5}}, {\mathbf{1}}, {\mathbf{2}}) \oplus ({\mathbf{1}}, {\mathbf{1}}, {\mathbf{2}}) \oplus ({\mathbf{10}}, {\mathbf{1}}, {\mathbf{1}})$
 %%%%%%%%%%%%%%%%%%%%%%%%%%%%%%%% 
  \\ \hline 118 &  $\mathfrak{su}(5)^{\oplus 2} \oplus \mathfrak{su}(2)$ & $({\mathbf{10}}, {\mathbf{1}}, {\mathbf{1}})^{\oplus 2} \oplus ({\mathbf{1}}, \overline{\mathbf{5}}, {\mathbf{1}}) \oplus ({\mathbf{1}}, {\mathbf{10}}, {\mathbf{1}}) \oplus (\overline{\mathbf{5}}, {\mathbf{1}}, {\mathbf{2}}) \oplus ({\mathbf{1}}, {\mathbf{1}}, {\mathbf{3}})$
 %%%%%%%%%%%%%%%%%%%%%%%%%%%%%%%% 
  \\ \hline 119 &  $\mathfrak{su}(5)^{\oplus 2} \oplus \mathfrak{su}(2)$ & $(\overline{\mathbf{5}}, {\mathbf{1}}, {\mathbf{1}})^{\oplus 2} \oplus ({\mathbf{10}}, {\mathbf{1}}, {\mathbf{1}})^{\oplus 2} \oplus ({\mathbf{1}}, \overline{\mathbf{5}}, {\mathbf{1}}) \oplus ({\mathbf{1}}, {\mathbf{10}}, {\mathbf{1}}) \oplus ({\mathbf{1}}, {\mathbf{1}}, {\mathbf{2}}) \oplus ({\mathbf{1}}, {\mathbf{1}}, {\mathbf{1}})$
 %%%%%%%%%%%%%%%%%%%%%%%%%%%%%%%% 
  \\ \hline 120 &  $\mathfrak{su}(5)^{\oplus 2} \oplus \mathfrak{su}(2)$ & $(\overline{\mathbf{5}}, {\mathbf{1}}, {\mathbf{1}})^{\oplus 2} \oplus ({\mathbf{10}}, {\mathbf{1}}, {\mathbf{1}})^{\oplus 2} \oplus ({\mathbf{1}}, \overline{\mathbf{5}}, {\mathbf{1}}) \oplus ({\mathbf{1}}, {\mathbf{10}}, {\mathbf{1}}) \oplus ({\mathbf{1}}, {\mathbf{1}}, {\mathbf{3}})$
 %%%%%%%%%%%%%%%%%%%%%%%%%%%%%%%% 
  \\
\hline 
\end{tabular}
\caption{Continued}
\makeatletter\onecolumngrid@pop\makeatother\end{table*}

\renewcommand{\arraystretch}{1.5} 
\setcounter{table}{1}
\makeatletter\onecolumngrid@push\makeatother\begin{table*}
\fontsize{8}{9.2}\selectfont
\begin{tabular}{|c|r|l|} \hline
\multicolumn{3}{|c|}{\small Non-Maximal and Non-Minimal algebras} \\ \hline
& {\small Algebra}&\small Fermion representations corresponding to \
$\beta$ \\ \hline 
161 &  $\mathfrak{su}(5) \oplus \mathfrak{so}(10) \oplus \mathfrak{su}(2)^{\oplus 2}$ & $({\mathbf{10}}, {\mathbf{1}}, {\mathbf{1}}, {\mathbf{1}})^{\oplus 2} \oplus ({\mathbf{1}}, {\mathbf{16}}, {\mathbf{1}}, {\mathbf{1}}) \oplus (\overline{\mathbf{5}}, {\mathbf{1}}, {\mathbf{2}}, {\mathbf{1}}) \oplus ({\mathbf{1}}, {\mathbf{1}}, {\mathbf{1}}, {\mathbf{2}})$
 %%%%%%%%%%%%%%%%%%%%%%%%%%%%%%%% 
  \\ \hline 162 &  $\mathfrak{su}(5) \oplus \mathfrak{su}(2)^{\oplus 3}$ & $(\overline{\mathbf{5}}, {\mathbf{1}}, {\mathbf{1}}, {\mathbf{1}}) \oplus ({\mathbf{10}}, {\mathbf{1}}, {\mathbf{1}}, {\mathbf{1}}) \oplus (\overline{\mathbf{5}}, {\mathbf{2}}, {\mathbf{1}}, {\mathbf{1}}) \oplus ({\mathbf{10}}, {\mathbf{1}}, {\mathbf{2}}, {\mathbf{1}}) \oplus ({\mathbf{1}}, {\mathbf{1}}, {\mathbf{1}}, {\mathbf{2}}) \oplus ({\mathbf{1}}, {\mathbf{1}}, {\mathbf{1}}, {\mathbf{1}})$
 %%%%%%%%%%%%%%%%%%%%%%%%%%%%%%%% 
  \\ \hline 163 &  $\mathfrak{su}(5) \oplus \mathfrak{su}(2)^{\oplus 3}$ & $(\overline{\mathbf{5}}, {\mathbf{1}}, {\mathbf{1}}, {\mathbf{1}}) \oplus ({\mathbf{10}}, {\mathbf{1}}, {\mathbf{1}}, {\mathbf{1}}) \oplus (\overline{\mathbf{5}}, {\mathbf{2}}, {\mathbf{1}}, {\mathbf{1}}) \oplus ({\mathbf{10}}, {\mathbf{1}}, {\mathbf{2}}, {\mathbf{1}}) \oplus ({\mathbf{1}}, {\mathbf{1}}, {\mathbf{1}}, {\mathbf{3}})$
 %%%%%%%%%%%%%%%%%%%%%%%%%%%%%%%% 
  \\
\hline 
\end{tabular}
\caption{Continued}
\makeatletter\onecolumngrid@pop\makeatother\end{table*}

\renewcommand{\arraystretch}{1.5} 
\setcounter{table}{1}
\makeatletter\onecolumngrid@push\makeatother\begin{table*}
\fontsize{8}{9.2}\selectfont
\begin{tabular}{|c|r|l|} \hline
\multicolumn{3}{|c|}{\small Non-Maximal and Non-Minimal algebras} \\ \hline
& {\small Algebra}&\small Fermion representations corresponding to \
$\beta$ \\ \hline 
204 &  $\mathfrak{su}(4) \oplus \mathfrak{su}(2)^{\oplus 4}$ & $({\mathbf{4}}, {\mathbf{3}}, {\mathbf{2}}, {\mathbf{1}}, {\mathbf{1}}) \oplus (\overline{\mathbf{4}}, {\mathbf{1}}, {\mathbf{1}}, {\mathbf{2}}, {\mathbf{1}})^{\oplus 2} \oplus (\overline{\mathbf{4}}, {\mathbf{1}}, {\mathbf{1}}, {\mathbf{1}}, {\mathbf{2}})$
 %%%%%%%%%%%%%%%%%%%%%%%%%%%%%%%% 
  \\ \hline 205 &  $\mathfrak{su}(4) \oplus \mathfrak{su}(2)^{\oplus 4}$ & $({\mathbf{4}}, {\mathbf{2}}, {\mathbf{2}}, {\mathbf{1}}, {\mathbf{1}}) \oplus (\overline{\mathbf{4}}, {\mathbf{1}}, {\mathbf{1}}, {\mathbf{2}}, {\mathbf{1}})^{\oplus 3} \oplus ({\mathbf{4}}, {\mathbf{1}}, {\mathbf{1}}, {\mathbf{1}}, {\mathbf{2}})$
 %%%%%%%%%%%%%%%%%%%%%%%%%%%%%%%% 
  \\ \hline 206 &  $\mathfrak{su}(4) \oplus \mathfrak{su}(2)^{\oplus 4}$ & $({\mathbf{4}}, {\mathbf{2}}, {\mathbf{2}}, {\mathbf{1}}, {\mathbf{1}}) \oplus (\overline{\mathbf{4}}, {\mathbf{1}}, {\mathbf{1}}, {\mathbf{2}}, {\mathbf{1}})^{\oplus 3} \oplus ({\mathbf{4}}, {\mathbf{1}}, {\mathbf{1}}, {\mathbf{1}}, {\mathbf{2}})$
 %%%%%%%%%%%%%%%%%%%%%%%%%%%%%%%% 
  \\ \hline 207 &  $\mathfrak{su}(4) \oplus \mathfrak{su}(2)^{\oplus 4}$ & $({\mathbf{4}}, {\mathbf{2}}, {\mathbf{2}}, {\mathbf{1}}, {\mathbf{1}}) \oplus ({\mathbf{4}}, {\mathbf{2}}, {\mathbf{1}}, {\mathbf{1}}, {\mathbf{1}}) \oplus (\overline{\mathbf{4}}, {\mathbf{1}}, {\mathbf{1}}, {\mathbf{2}}, {\mathbf{1}})^{\oplus 2} \oplus (\overline{\mathbf{4}}, {\mathbf{1}}, {\mathbf{1}}, {\mathbf{1}}, {\mathbf{2}})$
 %%%%%%%%%%%%%%%%%%%%%%%%%%%%%%%% 
  \\
\hline 
\end{tabular}
\caption{Continued}
\makeatletter\onecolumngrid@pop\makeatother\end{table*}

\renewcommand{\arraystretch}{1.5} 
\setcounter{table}{1}
\makeatletter\onecolumngrid@push\makeatother\begin{table*}
\fontsize{8}{9.2}\selectfont
\begin{tabular}{|c|r|l|} \hline
\multicolumn{3}{|c|}{\small Non-Maximal and Non-Minimal algebras} \\ \hline
& {\small Algebra}&\small Fermion representations corresponding to \
$\beta$ \\ \hline 
246 &  $\mathfrak{su}(4) \oplus \mathfrak{so}(10) \oplus \mathfrak{su}(2)^{\oplus 4}$ & $({\mathbf{1}}, {\mathbf{16}}, {\mathbf{1}}, {\mathbf{1}}, {\mathbf{1}}, {\mathbf{1}}) \oplus ({\mathbf{4}}, {\mathbf{1}}, {\mathbf{2}}, {\mathbf{2}}, {\mathbf{1}}, {\mathbf{1}}) \oplus (\overline{\mathbf{4}}, {\mathbf{1}}, {\mathbf{1}}, {\mathbf{1}}, {\mathbf{2}}, {\mathbf{1}}) \oplus (\overline{\mathbf{4}}, {\mathbf{1}}, {\mathbf{1}}, {\mathbf{1}}, {\mathbf{1}}, {\mathbf{2}})$
 %%%%%%%%%%%%%%%%%%%%%%%%%%%%%%%% 
  \\ \hline 247 &  $\mathfrak{su}(4) \oplus \mathfrak{so}(10) \oplus \mathfrak{su}(2)^{\oplus 4}$ & $({\mathbf{1}}, {\mathbf{16}}, {\mathbf{1}}, {\mathbf{1}}, {\mathbf{1}}, {\mathbf{1}}) \oplus ({\mathbf{4}}, {\mathbf{1}}, {\mathbf{2}}, {\mathbf{2}}, {\mathbf{1}}, {\mathbf{1}}) \oplus (\overline{\mathbf{4}}, {\mathbf{1}}, {\mathbf{1}}, {\mathbf{1}}, {\mathbf{2}}, {\mathbf{1}}) \oplus (\overline{\mathbf{4}}, {\mathbf{1}}, {\mathbf{1}}, {\mathbf{1}}, {\mathbf{1}}, {\mathbf{2}})$
 %%%%%%%%%%%%%%%%%%%%%%%%%%%%%%%% 
  \\ \hline 248 &  $\mathfrak{su}(4) \oplus \mathfrak{so}(10) \oplus \mathfrak{su}(2)^{\oplus 4}$ & $({\mathbf{1}}, {\mathbf{16}}, {\mathbf{1}}, {\mathbf{1}}, {\mathbf{1}}, {\mathbf{1}}) \oplus (\overline{\mathbf{4}}, {\mathbf{1}}, {\mathbf{2}}, {\mathbf{1}}, {\mathbf{1}}, {\mathbf{1}}) \oplus (\overline{\mathbf{4}}, {\mathbf{1}}, {\mathbf{1}}, {\mathbf{2}}, {\mathbf{1}}, {\mathbf{1}}) \oplus ({\mathbf{4}}, {\mathbf{1}}, {\mathbf{1}}, {\mathbf{1}}, {\mathbf{2}}, {\mathbf{1}}) \oplus ({\mathbf{4}}, {\mathbf{1}}, {\mathbf{1}}, {\mathbf{1}}, {\mathbf{1}}, {\mathbf{2}})$
 %%%%%%%%%%%%%%%%%%%%%%%%%%%%%%%% 
  \\
\hline 
\end{tabular}
\caption{Continued}
\makeatletter\onecolumngrid@pop\makeatother\end{table*}

\renewcommand{\arraystretch}{1.5} 
\setcounter{table}{1}
\makeatletter\onecolumngrid@push\makeatother\begin{table*}
\fontsize{8}{9.2}\selectfont
\begin{tabular}{|c|r|l|} \hline
\multicolumn{3}{|c|}{\small Non-Maximal and Non-Minimal algebras} \\ \hline
& {\small Algebra}&\small Fermion representations corresponding to \
$\beta$ \\ \hline 
289 &  $\mathfrak{su}(4)^{\oplus 2} \oplus \mathfrak{su}(2)^{\oplus 4}$ & $(\overline{\mathbf{4}}, {\mathbf{1}}, {\mathbf{2}}, {\mathbf{1}}, {\mathbf{1}}, {\mathbf{1}})^{\oplus 2} \oplus ({\mathbf{4}}, {\mathbf{1}}, {\mathbf{1}}, {\mathbf{2}}, {\mathbf{1}}, {\mathbf{1}})^{\oplus 2} \oplus ({\mathbf{1}}, \overline{\mathbf{4}}, {\mathbf{1}}, {\mathbf{1}}, {\mathbf{2}}, {\mathbf{1}}) \oplus ({\mathbf{1}}, {\mathbf{4}}, {\mathbf{1}}, {\mathbf{1}}, {\mathbf{1}}, {\mathbf{2}})$
 %%%%%%%%%%%%%%%%%%%%%%%%%%%%%%%% 
  \\ \hline 290 &  $\mathfrak{su}(4)^{\oplus 3} \oplus \mathfrak{su}(2)^{\oplus 3}$ & $(\overline{\mathbf{4}}, {\mathbf{1}}, {\mathbf{1}}, {\mathbf{2}}, {\mathbf{1}}, {\mathbf{1}}) \oplus ({\mathbf{1}}, \overline{\mathbf{4}}, {\mathbf{1}}, {\mathbf{2}}, {\mathbf{1}}, {\mathbf{1}}) \oplus ({\mathbf{1}}, {\mathbf{1}}, \overline{\mathbf{4}}, {\mathbf{2}}, {\mathbf{1}}, {\mathbf{1}}) \oplus ({\mathbf{4}}, {\mathbf{1}}, {\mathbf{1}}, {\mathbf{1}}, {\mathbf{2}}, {\mathbf{1}}) \oplus ({\mathbf{1}}, {\mathbf{4}}, {\mathbf{1}}, {\mathbf{1}}, {\mathbf{2}}, {\mathbf{1}}) $
\\ & & $ \oplus ({\mathbf{1}}, {\mathbf{1}}, {\mathbf{4}}, {\mathbf{1}}, {\mathbf{1}}, {\mathbf{2}})$
  %%%%%%%%%%%%%%%%%%%%%%%%%%%%%%%% 
  \\ \hline 291 &  $\mathfrak{su}(4)^{\oplus 3} \oplus \mathfrak{su}(2)^{\oplus 3}$ & $(\overline{\mathbf{4}}, {\mathbf{1}}, {\mathbf{1}}, {\mathbf{2}}, {\mathbf{1}}, {\mathbf{1}}) \oplus ({\mathbf{1}}, \overline{\mathbf{4}}, {\mathbf{1}}, {\mathbf{2}}, {\mathbf{1}}, {\mathbf{1}}) \oplus ({\mathbf{1}}, {\mathbf{1}}, \overline{\mathbf{4}}, {\mathbf{2}}, {\mathbf{1}}, {\mathbf{1}}) \oplus ({\mathbf{4}}, {\mathbf{1}}, {\mathbf{1}}, {\mathbf{1}}, {\mathbf{2}}, {\mathbf{1}}) \oplus ({\mathbf{1}}, {\mathbf{4}}, {\mathbf{1}}, {\mathbf{1}}, {\mathbf{2}}, {\mathbf{1}}) $
\\
\hline 
\end{tabular}
\caption{Continued}
\makeatletter\onecolumngrid@pop\makeatother\end{table*}

\renewcommand{\arraystretch}{1.5} 
\setcounter{table}{1}
\makeatletter\onecolumngrid@push\makeatother\begin{table*}
\fontsize{8}{9.2}\selectfont
\begin{tabular}{|c|r|l|} \hline
\multicolumn{3}{|c|}{\small Non-Maximal and Non-Minimal algebras} \\ \hline
& {\small Algebra}&\small Fermion representations corresponding to \
$\beta$ \\ \hline 
324 &  $\mathfrak{su}(4)^{\oplus 2} \oplus \mathfrak{su}(2)^{\oplus 5}$ & $(\overline{\mathbf{4}}, {\mathbf{1}}, {\mathbf{2}}, {\mathbf{1}}, {\mathbf{1}}, {\mathbf{1}}, {\mathbf{1}}) \oplus ({\mathbf{1}}, \overline{\mathbf{4}}, {\mathbf{2}}, {\mathbf{1}}, {\mathbf{1}}, {\mathbf{1}}, {\mathbf{1}}) \oplus (\overline{\mathbf{4}}, {\mathbf{1}}, {\mathbf{1}}, {\mathbf{2}}, {\mathbf{1}}, {\mathbf{1}}, {\mathbf{1}}) \oplus ({\mathbf{4}}, {\mathbf{1}}, {\mathbf{1}}, {\mathbf{1}}, {\mathbf{2}}, {\mathbf{1}}, {\mathbf{1}}) \oplus ({\mathbf{4}}, {\mathbf{1}}, {\mathbf{1}}, {\mathbf{1}}, {\mathbf{1}}, {\mathbf{2}}, {\mathbf{1}}) $
\\ & & $ \oplus ({\mathbf{1}}, {\mathbf{4}}, {\mathbf{1}}, {\mathbf{1}}, {\mathbf{1}}, {\mathbf{1}}, {\mathbf{2}})$
  %%%%%%%%%%%%%%%%%%%%%%%%%%%%%%%% 
  \\ \hline 325 &  $\mathfrak{su}(4)^{\oplus 2} \oplus \mathfrak{su}(2)^{\oplus 5}$ & $(\overline{\mathbf{4}}, {\mathbf{1}}, {\mathbf{2}}, {\mathbf{1}}, {\mathbf{1}}, {\mathbf{1}}, {\mathbf{1}}) \oplus ({\mathbf{1}}, \overline{\mathbf{4}}, {\mathbf{2}}, {\mathbf{1}}, {\mathbf{1}}, {\mathbf{1}}, {\mathbf{1}}) \oplus (\overline{\mathbf{4}}, {\mathbf{1}}, {\mathbf{1}}, {\mathbf{2}}, {\mathbf{1}}, {\mathbf{1}}, {\mathbf{1}}) \oplus ({\mathbf{4}}, {\mathbf{1}}, {\mathbf{1}}, {\mathbf{1}}, {\mathbf{2}}, {\mathbf{1}}, {\mathbf{1}}) \oplus ({\mathbf{4}}, {\mathbf{1}}, {\mathbf{1}}, {\mathbf{1}}, {\mathbf{1}}, {\mathbf{2}}, {\mathbf{1}}) $
\\ & & $ \oplus ({\mathbf{1}}, {\mathbf{4}}, {\mathbf{1}}, {\mathbf{1}}, {\mathbf{1}}, {\mathbf{1}}, {\mathbf{2}})$
  %%%%%%%%%%%%%%%%%%%%%%%%%%%%%%%% 
  \\
\hline 
\end{tabular}
\caption{Continued}
\makeatletter\onecolumngrid@pop\makeatother\end{table*}

%%%%%%%%%%%%%%%%%%%%%%

\end{document}